\newcommand{\ddx}{{\partial \over \partial x}}
\newcommand{\ddt}{{\partial \over \partial t}}
\newcommand{\eps}{\epsilon}
\documentstyle[preprint,aps]{revtex}
\begin{document}
\draft

\title{Density Waves in the Flows of Granular Media}
\author{Jysoo Lee}
\address{HLRZ-KFA J\"{u}lich, Postfach 1913, W-5170 J\"{u}lich, Germany}
\date{\today}
\maketitle

\begin{abstract}
We study density waves in the flows of granular particles through
vertical tubes and hoppers using both analytic methods and molecular
dynamics (MD) simulations. We construct equations of motion for quasi
one-dimensional systems. The equations, combined with the Bagnold's
law for friction, are used to describe the time evolutions of the
density and the velocity fields for narrow tubes and hoppers. The
solutions of the equations can have two types of density waves,
kinetic and dynamic. For tubes, we can show the existence of kinetic
waves, and obtain the condition for dynamic waves for tubes from the
equations. For hoppers, we obtain the solutions of the equations up to
the first order of the opening angle, which also show the existence of
kinetic waves. We reproduce density waves in the MD simulations for
tubes.  The waves are believed to be kinetic based on a few evidences,
including a well defined flux-density curve. In MD simulations of
flows in hoppers, we find density waves, which are also believed to be
kinetic.
\end{abstract}


\section{Introduction}

Systems of granular particles (e.g. sands) exhibit many interesting
phenomena \cite{s84,cc90,jn92}. The formations of a spontaneous heap
\cite{er89,ldf89,cdr92,l92} and convection cells
\cite{r76,s88,zs91,t92,gh92} under vibration, and the segregation of
particles \cite{w76,hw86,rsps87,d90,jm92,drc93} are just a few
examples. These phenomena are consequences of unusual, and often
complex, dynamical responses of the system. Considering the complexity
of the dynamics, one is tempted to first have comprehensive
understanding of the dynamics of granular media in relatively simple
geometries, and proceed to more complicated situations. Even for the
simple geometries like shear cells, vertical tubes and hoppers,
granular media still show complex dynamics. For example, granular
particles in a shear cell show not only non-Newtonian behaviors
\cite{b54,ss84,c89,bss90}, but also stick-slip motions \cite{tg91} as well
as density waves \cite{hl90,ss92}. Here, we study the flows of
granular particles through vertical tubes and hoppers, whose
geometries are as simple as shear cells, but are less well studied. In
granular flows through vertical tubes, P\"{o}schel \cite{p92} found
that the particles do not flow uniformly, but form regions of high
density which travel with velocity different from that of the center
of mass. The mechanism for the traveling density pattern or the
density wave is, however, not clearly understood. The density waves
are also found in outflows through hoppers \cite{cp67,bbfj89}.
Especially, Baxter {\it et al} \cite{bbfj89} show that the velocities
of the density waves are dependent on the opening angle of hoppers,
and even their directions can be changed. Furthermore, they find the
density waves only when they mix some amount of rough sands with
smooth sands. In MD simulations of hopper, the density field is found
out to be non-uniform \cite{r92}. Again, the mechanism for the density
wave is not clearly understood.

We first construct equations of motion for quasi one-dimensional
systems. With the Bagnold's law for friction
\cite{b54}, the equations are used to describe the time evolutions of
the density and the velocity fields for narrow tubes and hoppers. The
solutions of the equations can have two types of density waves,
kinetic and dynamic. The dynamic waves are similar to sound waves,
while kinetic waves are due to the kinematics of the equations
\cite{lw55}. For tubes, we can show the existence of kinetic waves, and
obtain the condition for dynamic waves by a linear stability analysis.
We also study the effects of static friction, and find that it changes
the qualitative behaviors of the flows. For hoppers, we obtain the
solutions of the equations up to the first order of the opening angle,
and also show the existence of kinetic waves. Next, we study the
systems using molecular dynamics (MD) simulations. For tubes, we
reproduce density waves, and we find a few strong evidences that the
waves are of kinetic nature, which includes a well defined
flux-density curve and the density dependence of the velocities of the
waves. We also study the existence of dynamic waves by changing the
inelasticity of the particles, and do not find dynamics waves. In MD
simulations of flows in hoppers, we find density waves, which are
believed to be kinetic waves due to the strong correlation between the
density and the velocity fields. For hoppers with periodic boundary
conditions, we find additional density waves, which eventually
dominates. We find a few evidences that the additional waves are of
dynamic nature. We, however, do not find dynamic waves in hopper with
open boundary conditions. The results obtained by these theoretical
and numerical studies are compared with the experiments.

This paper is organized as follows. In Sec.\ \ref{sec:theory}, we
present theoretical results. Detailed discussions on the definitions
and properties of dynamic and kinetic waves are given in Sec.\
\ref{sec:dvsk}. The kinetic and dynamic waves in tubes are discussed
in Sec.\ \ref{sec:thtube} and Sec.\ \ref{sec:dynamic}, respectively.
Also, kinetic waves in hoppers are discussed in Sec.\ \ref{sec:thhop}.
We present the numerical results in Sec.\ \ref{sec:numerical}. First,
we define the interactions between the particles in Sec.\
\ref{sec:inter}. The results for kinetic and dynamic waves in tubes
are given in Sec.\ \ref{sec:nutube}, and the results for hoppers are
discussed in Sec.\ \ref{sec:nuhop}. Brief summary as well as the
limitations of this work are given in Sec.\ \ref{sec:dis}.

\section{Theoretical Approach}
\label{sec:theory}

\subsection{Dynamic and kinetic waves}
\label{sec:dvsk}

Here, we construct equations of motion for flows in one-dimension,
which will later be used to describe the time evolutions of fields in
vertical tubes and hoppers. Let us define $\rho (x,t)$ and $v(x,t)$ be
the density and the velocity at position $x$ and time $t$,
respectively. If we require the mass of the system to be conserved,
\begin{mathletters}
\label{eq:motion}
\begin{equation}
\ddt \rho + \ddx (\rho v) = 0.
\label{eq:mass}
\end{equation}
Other equation comes from the momentum conservation,
\begin{equation}
\rho \ddt v + \rho v \ddx v = F(x,t),
\label{eq:momentum}
\end{equation}
\end{mathletters}
where $F(x,t)dx$ is the total force acting on the mass within
$[x,x+dx]$ at time $t$.

There are two distinct mechanisms by which density patterns can
travel. One is due to the instability of uniform-density flows to
density fluctuations (``dynamic wave''), and the other is of kinetic
origin (``kinetic wave'') \cite{lw55}. We now discuss both types of
waves in detail, paying particular attention to their differences. In
order to make the discussion more concrete, we use a specific form of
the force
\begin{equation}
\label{eq:force}
F(x,t) = \rho - \ddx \rho - \rho ^{\nu} v^{\mu},
\end{equation}
where $\mu$ and $\nu$ are constants. The form of Eq.\
(\ref{eq:force}), motivated from a granular flow in a tube, is chosen
such that the discussion becomes more clear. The followings are,
however, applicable for any form of the force.

We first consider the dynamic wave. The equations of motion Eqs.\
(\ref{eq:motion}) with the force given by Eq.\ (\ref{eq:force}) have a
solution of uniform density $\rho _o$,
\begin{eqnarray}
\label{eq:uniform}
\rho (x,t) &&= \rho _o, \cr
v(x,t) &&= \rho _o^{(1-\nu)/\mu}.
\end{eqnarray}
We study the stability of the solution Eq.\ (\ref{eq:uniform}) by a
linear stability analysis. For simplicity, we consider the case of
$\rho _o = 1$. The trial solution for the time evolution of a
perturbation of Eq.\ (\ref{eq:uniform}) is
\begin{eqnarray}
\label{eq:perturbe}
\rho (x,t) &&= 1 + \eps _{\rm \rho} \exp [i(kx-\omega t)] \cr
v(x,t) &&= 1 + \eps _{\rm v} \exp [i(kx-\omega t)],
\end{eqnarray}
where $\eps _{\rm \rho}$ and $\eps _{\rm v}$ are constants. We want to
study the time evolution in a linear approximation, where we consider
terms up to the first order of $\eps$. The analysis is valid only if a
perturbation term is much smaller than unity. Substituting Eq.\
(\ref{eq:perturbe}) into Eqs.\ (\ref{eq:motion}), and discard terms
higher order than $\eps$, we obtain
\begin{eqnarray}
\label{eq:linear}
(\omega - k) \eps _{\rm \rho} -  k \eps _{\rm v} &&= 0 \cr
[k + i(1 - \nu )] \eps _{\rm \rho} + [(k - \omega ) - i \mu] \eps _{\rm
v} &&= 0,
\end{eqnarray}
whose solution is given by
\begin{equation}
\label{eq:wk}
\omega = {2k - i\mu \pm \sqrt{4k[k+i(1-\nu )] - \mu ^{\rm 2}} \over
2}.
\end{equation}
The density fluctuation travels with velocity $Re(\omega ) / k$, where
$Re(\omega )$ is the real part of $\omega $. If $Im(\omega )$ (the
imaginary part of $\omega $) is positive, the perturbation grows with
time, makes uniform-density flow unstable. This mechanism for creating
density waves, which is based on the instability of the equations, is
called dynamic waves.

We now discuss the mechanism for kinetic waves. For an excellent
introduction to the subject, please see Ref.\ \cite{lw55}. Consider a
case that the system is divided into two uniform-density flow regions
with densities $\rho _a$ and $\rho _b$, and the velocities of each
region is determined by Eq.\ (\ref{eq:uniform}) (Fig.\
\ref{fig:inter}). The equations of motion Eqs.\ (\ref{eq:motion}) imply
that density and velocity field ($\rho (x,t)$ and $v(x,t)$) inside
each domains do not change.  The system evolve only by moving the
interface.  Let $U$ be the velocity of the interface. We also set the
initial position of the interface to be $0$ without losing generality.
Then, the position of the interface at time $t$ is $Ut$. At a given
time, we choose an interval $[-\eps ,\eps ]$ which includes the
interface. We then integrate Eq.\ (\ref{eq:mass}) in the interval
\begin{eqnarray}
\label{eq:kvstep}
\int_{-\eps}^{\eps} \ddt \rho + \ddx (\rho v) = && \ddt [(\eps + Ut)\rho
_a + (\eps - Ut)\rho _b] + \rho v |_{-\eps}^{\eps} \cr = && (\rho
a - \rho _b)U + j_b - j_a = 0,
\end{eqnarray}
where the flux $j$ is defined to be $\rho v$. Therefore, the
interface, or any density fluctuations, travels with velocity
\begin{equation}
\label{eq:kvel}
U = {j_a - j_b \over \rho _a - \rho _b}~=~{\Delta j \over \Delta
\rho}.
\end{equation}
Since the above mechanism for the travel of density fluctuations is a
consequence of a kinetic equation (mass conservation, Eq.\
(\ref{eq:mass})), it is called kinetic wave. Unlike dynamics waves,
the fluctuations in kinetic waves can not be amplified, the system can
only reorganize existing fluctuations. The fluctuations are created by
initial conditions or by noises in the dynamics.  Furthermore, kinetic
waves, in most cases, decay into uniform flows \cite{lw55}.

In general, flows in $1$-d have both kinetic and dynamic waves. If a
system is stable under density fluctuations, existing dynamic waves
will decay exponentially, and only kinetic waves can survive. On the
other hand, if the system is unstable, the amplitudes of the dynamic
waves will grow, and they eventually dominate over kinetic waves. In
order to illustrate this point, we integrate numerically Eqs.\
(\ref{eq:motion}) to get $\rho (x,t)$ and $v(x,t)$, where the force is
given by Eq.\ (\ref{eq:force}). The integration is done using the
staggered leapfrog method \cite{pftv86} with the time step $0.001$.
We also use a periodic boundary condition. The initial conditions are
that $\rho (x,0)$ is a short pulse of higher density (1 \%) on the top
of background density of unity, and $v(x,0)$ is given by Eq.\
(\ref{eq:uniform}). We first study the case of $\mu = 2, \nu = -1.$
The behavior of the dynamic waves is studied by a linear stability
analysis. Equation (\ref{eq:wk}) gives the dependence of $\omega$ on
$k$ to be $2k, -2i$. There is one marginally stable mode of velocity
$2$, and one stable mode of velocity $0$. On the other hand, the flux
is given as
\begin{equation}
\label{eq:kvel2}
j \equiv \rho v = \rho ^{(\mu-\nu+1)/\mu},
\end{equation}
and the velocity of kinetic wave, determined by Eq.\ (\ref{eq:kvel}),
is approximately $2$. Since the system is not unstable to density
fluctuations, the kinetic waves, not greatly affected by the dynamic
waves, will survive for a long time (Fig.\
\ref{fig:dvsk}(a)). We now study the case of $\mu =-2, \nu = -1$.
Following the same analysis, $\omega (k)$ is determined to be $2k +
2i$ and $0$. One unstable mode of velocity $2$ and one marginally
stable mode of velocity $0$. The velocity of the kinetic wave is
approximately $0$. The dynamic wave of increasing amplitude (that of
the velocity $2$) is dominating over the kinetic wave (Fig.\
\ref{fig:dvsk}(b)).

\subsection{Density waves in vertical tubes}

\subsubsection{Kinetic waves in vertical tubes}
\label{sec:thtube}

We consider vertical tubes of narrow width, which are filled with
granular material. The particles will flow down due to gravity. This
motion will induce the forces between the particles as well as the
friction forces by the inner walls of the tubes. We assume that these
forces follow the Bagnold's law \cite{b54}. The law, first found in a
rapid granular flow, implies that the components of the stress tensor
$\tau _{ij}$ are proportional to the square of the shear rate
$\dot{\gamma }$,
\begin{equation}
\label{eq:bagnold}
\tau _{ij} = \rho _B D^{2} f_{ij}(p) \dot{\gamma }^{2}.
\end{equation}
Here, $\rho _{B}$ is the density of the material which forms the
particles, and $p$ is the volume fraction of the particles. The
density $\rho $ is, therefore, $\rho _{B}$ times $p$.  Also, $D$ is
the average diameter of the particles, $f_{ij}(p)$ is a system
dependent function. Bagnold predicted the shear rate dependence in
rapid flows using a simple argument, which later confirmed by more
elaborate calculations as well as experiments
\cite{cc90,jn92,ss84}. In this paper, we will use the law except
the case that the velocities of the particles are considered to be
very small.

Let $x$-axis ($y$-axis) be the direction parallel (perpendicular) to
the tube, where the positive $x$-direction is chosen to be upward. We
consider the forces acting on the granular material contained in
$[x,x+dx]$. Since we study only narrow tubes, we assume the material
is homogeneous along the $y$ direction. There is a gravitational force
given by $-\rho g W dx,$ where $g$ is the gravitational acceleration,
and $W$ the width of the tube.  There also is a friction by the wall,
which is $-\tau _{xy} dx$, and pressure by other particles $-d \tau
_{xx} /dx~D dx$. The total force $F(x,t)$ becomes
\begin{eqnarray}
\label{eq:reforce}
F(x,t) = && -\rho _B p g W - sign(v) \rho _B D^{2} f_{xy}(p)
({v \over D})^{2} - D \ddx [\rho _B D^{2} f_{xx}(p) ({v
\over D})^{2}] \cr = && -\rho _B p g W - sign(v) \rho _B
f_{xy}(p) v^{2} - D \ddx [\rho _B f_{xx}(p) v^{2}],
\end{eqnarray}
where $sign(v)$ is the sign of $v$, and we assume the width of shear
layer to be of order of $D$.

Having found the force, we now proceed to study kinetic waves in the
system. We first consider the case of {\it no} static friction. The
steady-state velocity $v_{s}(p)$ of a uniform flow is obtained from
the condition that the total force acting on the granular material is
zero. Since we consider uniform density flows, the force free
condition gives
\begin{equation}
\label{eq:terminal}
v_{s}(p) = -\sqrt{pgW/f_{xy}(p)}.
\end{equation}
The flux $j(p)$ is given by $\rho _B p v_{s}(p)$, then the velocity of
a small density fluctuation $U(p)$ becomes
\begin{equation}
\label{eq:rekvel}
U(p) = {\Delta j \over \Delta \rho} \simeq {dj(p) \over d\rho } = {1
\over 2} v_{s}(p) {3f_{xy}(p) - pdf_{xy}(p)/dp \over f_{xy}(p)}.
\end{equation}
Therefore, the form of the force Eq.\ (\ref{eq:reforce}) allows
density fluctuations to travel with velocity $U(p)$. The condition for
the existence of kinetic waves is the balance between the forces, and
the waves are not very sensitively dependent on the exact form of the
forces. Therefore, they are very likely to be seen in the experiments.

We now consider the effects of static friction on the kinetic waves.
By static friction, we mean one has to apply a finite force in order
to break contacts between surfaces. There are two contributions to the
normal stress on the wall. One is a static pressure $P_{s}$, which is
independent of the motion of the particles. The other is due to the
collisions of the particles on the wall $\tau _{yy}$, given by Eq.\
(\ref{eq:bagnold}). The total friction force is the minimum of $\rho g
W - \tau _{xy}$ and $\mu (\tau _{yy}+P_{s})$, where $\mu $ is the
friction coefficient. If the shear force (gravity and friction) is
smaller than $\mu$ times the normal force, the total force on the
particles is zero, and the particles form a stagnant zone of no
movement \cite{zc92}.

Depending on the parameters, there can be two behaviors of the system.
If the gravitational force $\rho g W$ is less than the friction due to
static pressure $\mu P_{s}$, the particles can not move, and the
system remains as a static column. If $\rho g W > \mu P_{s}$, the
particles will start move. The steady-state velocity $v_{s}^{\prime}
(p)$ for given packing fraction $p$ is,
\begin{equation}
\label{eq:rekvelmu}
v_{s}^{\prime} (p) = - \sqrt{ p g W - \mu P_{s} / \rho _{B} \over
f_{xy}(p)},
\end{equation}
which is very similar to Eq.\ (\ref{eq:terminal}). However, we have
additional constraint that the shear force should exceed $\mu$ times
the normal force, which gives
\begin{equation}
\label{eq:static}
\rho g W - \rho _{B} f_{xy}(p) v^{2} > \mu P_{s} + \mu \rho _{B}
f_{yy}(p) v^{2}.
\end{equation}
If we define $v_{c}(p)$ be the velocity that above equation becomes
equality,
\begin{equation}
v_{c}(p) = v_{s}^{\prime} (p) {1 \over 1 + \mu f_{yy}(p) / f_{xy}(p)} <
v_{s}^{\prime} (p).
\end{equation}
Since the inequality Eq.\ (\ref{eq:static}) is not satisfied for $v(p)
> v_{c}(p)$, the particles can not reach their steady state. Instead,
they forms a stagnant zone. If we assume that the particles are
stopped completely in the stagnant zone, the distances between the
zones ${\cal D}(p)$ are constant, which is the distance needed for the
particles to be accelerated to $v_{c}(p)$ from the stopped state. We
can calculate the distance ${\cal D}(p)$ as follows. The velocity of
the particles, obtained by integrating the Newton's equation with the
friction force of Eq.\ (\ref{eq:static}), becomes
\begin{equation}
\label{eq:vtp}
v(t,p) = v_{s}^{\prime} (p) \tanh [v_{s}^{\prime} (p) ( g - {\mu P_{s}
\over
\rho W}) t].
\end{equation}
The distance covered by the particles during the interval $[0, t_{c}]$
becomes ${\cal D}(p)$, where $t_{c}$ is given by $v(t_{c},p) =
v_{c}(p)$. Integrating Eq.\ (\ref{eq:vtp}), we obtain
\begin{equation}
\label{eq:tubesep}
D(p) = -{v_{s}^{\prime 2}(p) \over 2 (g - \mu P_{s} / \rho W)} \ln (1 -
\gamma ^{2}),
\end{equation}
where $\gamma = 1 / (1 + \mu f_{yy}(p) / f_{xy}(p))$. The distance
${\cal D}(p)$ increases as $\mu$ is decreased, and it diverges when
$\mu \to 0$.

We now summerize the results obtained in this section. If static
friction is not present, the particles in vertical tubes form density
waves travels with velocity $U(p)$. When we take into account static
friction, the system forms periodic stagnant zones with their
separations ${\cal D}(p)$.

\subsubsection{Dynamic waves in vertical tubes}

\label{sec:dynamic}
In this section, we study the stability of the uniform-density flows
through vertical tubes under density and velocity fluctuations. We
will follow the analysis presented in Sec.\ \ref{sec:dvsk}. We start
with the equations of motion Eqs.\ (\ref{eq:motion}) with the force
given by Eq.\ (\ref{eq:reforce}). Here, we do not consider static
friction. The uniform-density flow solution with the density $\rho
_{o}$ is
\begin{eqnarray}
\label{eq:tubeuni}
\rho (x,t) &&= \rho _{o} \equiv \rho _{B} p_{o} \cr
v(x,t) &&= v_{o} \equiv - \sqrt{p_{o} g W \over f_{xy} (p_{o})}.
\end{eqnarray}
We study the stability of the solution by studying the time evolution
of the perturbation from the uniform-density flow, given by
\begin{eqnarray}
\label{eq:tubepe}
p (x,t) &&= p _{o} + \eps _{p} \exp [i(kx-\omega t)]
\cr
v(x,t) &&= v_{o} + \eps _{v} \exp [i(kx-\omega t)].
\end{eqnarray}
Substituting Eqs.\ (\ref{eq:tubepe}) into Eq.\ (\ref{eq:motion}), and
consider only up to the first terms of $\eps$, we obtain
\begin{eqnarray}
\label{eq:tubeset}
(k v_{o} - \omega) \eps _{p} + p _{o} k \eps _{v} &&= 0 \cr [g -
{v_{o}^{2} \over W} ({df_{xx} \over dp} - i k D {df_{xy} \over dp})]
\eps _{p} + [i (k v_{o} - \omega ) - {2 v_{o} \over W} (f_{xy} - i k D
f_{xx})] \eps _{v} &&= 0.
\end{eqnarray}
Since the two equations in Eq.\ (\ref{eq:tubeset}) are valid for any
values of $\eps _{p}$ and $\eps _{v}$, the solution is
\begin{equation}
\label{eq:omega}
\Omega ^{2} - i {2 v_{o} \over p_{o} W} (f_{xy} - i k D f_{xx})
\Omega + i k [g - {v_{o}^{2} \over W}({df_{xy} \over dp} - i k D
{df_{xx} \over dp})] = 0,
\end{equation}
where $\Omega \equiv \omega - k v_{o}$. The stability of the uniform
solution is determined by the imaginary part of $\omega$.
Unfortunately, since $\omega$ depends on the exact form of the unknow
function $f_{ij}$, we can not determine the stability of the flow.
Similar stability analysis on shear cells \cite{ss92} and dissipative
gases \cite{gz93} shows that the system is unstable for small
coefficient of restitution $e$. Especially, in \cite{gz93}, this
instability is traced back to the fact that pressure can decrease as
the density is increased. Here, we can show that Eq.\ (\ref{eq:omega})
has an unstable mode if $df_{xx}/dp$ is sufficiently negative (smaller
pressure for larger density), where the exact criterion is a
complicated function of $k, f_{xy}$ and $f_{xx}$. The stability of the
flow is later checked by MD simulations, where we will study the
stability for various degrees of the inelasticity of the particles.

\subsection{Density waves in hoppers}
\label{sec:thhop}

Consider a hopper with the opening angle $2 \theta$, where the width
at position $x$ is given by $W(x) = W_{o} + 2 \tan \theta x$ (Fig.\
\ref{fig:hopper}(a)). Here, we consider only the positive ranges of
$x$. The equations of motion for flows in hoppers are slightly
different from those for tubes. Since the width $W$ is dependent on
$x$, the conservation of mass implies
\begin{equation}
\label{eq:hmass}
W \ddt \rho + \ddx (W \rho v) = 0,
\end{equation}
instead of Eq.\ (\ref{eq:mass}). Equation (\ref{eq:momentum}), which
is a consequence of momentum conservation, still holds for hoppers,
with a modified form of friction force. Since the sidewalls are
tilted, the friction per unit length along the walls have two
contributions. One is the component of $\tau _{xy}$ (friction force
for tubes) parallel to the walls $\tau _{xy} \cos \theta$. Forces on
the walls by the internal pressure $\tau _{xx}$ give additional
contribution $\tau _{xx} \sin \theta \cos \theta $, where $\sin
\theta$ is due to the cross-section for the collisions of particles
to the walls. (Fig.\ \ref{fig:hopper}(b)). Therefore, the total force
becomes,
\begin{equation}
\label{eq:hopforce}
F(x,t) = - \rho g + \rho _B v^{2} (f_{xy} \cos \theta + f_{xx} \sin
\theta \cos \theta ) - \rho _{B} D \ddx (f_{xx} v^{2}).
\end{equation}
Since we assume that the system is homogeneous in the horizontal
direction, we only consider hoppers of narrow width $W$. We also
assume $\theta \ll 1$, and the $\theta$ dependences of physical
quantities (density, velocity) are calculated up to the first order of
$\theta$.

We now study the steady state properties of hoppers. Imposing the
steady state condition on the mass conservation Eq.\ (\ref{eq:hmass})
gives
\begin{mathletters}
\label{eq:hop1}
\begin{equation}
\label{eq:hopmass1}
\ddx (W p v) = 0.
\end{equation}
Also, the momentum conservation Eq.\ (\ref{eq:momentum}) with the
force Eq.\ (\ref{eq:hopforce}), in a steady state, becomes
\begin{equation}
\label{eq:hopmom1}
p v \ddx v = - p g + {1 \over W} v^{2} (f_{xy} \cos \theta + f_{xx}
\sin \theta \cos \theta) - {D \over W} \ddx(f_{xx} v^{2}).
\end{equation}
\end{mathletters}
We want to know the density and velocity fields which satisfies Eq.\
(\ref{eq:hop1}). We start with a trial solution
\begin{eqnarray}
\label{eq:hoptry}
p(x) &&= p_{o} + A(x) \theta \cr
v(x) &&= v_{s} + B(x) \theta,
\end{eqnarray}
where $v_{s} = - \sqrt{p_o g W_{o} / f_{xy}(p_{o})}$, the steady state
velocity for tubes Eq.\ (\ref{eq:terminal}). Substituting Eq.\
(\ref{eq:hoptry}) into Eq.\ (\ref{eq:hop1}), and ignore terms of
higher order than $\theta$,
\begin{eqnarray}
\label{eq:hopsteady}
v_s W_o {dA \over dx} + p_o W_o {dB \over dx} && = - 2 p_o v_s
\cr D v_s^2 {df_{xx} \over dp} {dA \over dx} + (2 D v_s f_{xx} +
p_o v_s W_o) {dB \over dx} + g W_o A - 2 v_s f_{xy} B && = v_s^2
f_{xx} - 2 x g p_o.
\end{eqnarray}
Since $f_{xx}$ and $f_{xy}$ are also functions of $p(x)$, one has to
know the exact form of these functions to find the solutions of the
equation Eq.\ (\ref{eq:hopsteady}). Here, we assume that $f_{xx}(p)$
can be approximated to $f_{xx}^o p^m$ for the range of densities found
in a hopper. Similarly, $f_{xy} \simeq f_{xy}^o p.^m$  Since $\theta
\ll 1$, the changes of $A(x)$ and $B(x)$ to $x$ will be also slow. We
therefore consider variations up to the first order of $x$, that is,
$A(x) = A_o x + A_1$ and $B(x) = B_o x + B_1$.  Substituting $f_{xx},
f_{xy}$ and $A(x), B(x)$ into Eq.\ (\ref{eq:hopsteady}), we obtain
\begin{eqnarray}
\label{eq:hopsteady2}
v_s A_o + p_o B_o = &&- 2 {p_o v_s \over W_o} \cr 2 g p_o = && (m
v_s^2 f_{xy}^o p_o^{m-1} - g W_o) A_o + 2 v_w f_{xy}^o p_o^m B_o \cr
v_s p_o W_o B_o = && -g W_o A_1 + v_s^2 p_o^m f_{xx}^o + m f_{xy}^o
v_s^2 p_o^{m-1} A_1 \cr && + 2 f_{xy}^o v_s p_o^m B_1 - D f_{xx}^o v_s
p_o^m (2 B_o + {m v_s A_o \over p_o}).
\end{eqnarray}
There are only three conditions in Eq.\ (\ref{eq:hopsteady2}) for four
unknown variables. The other condition comes from the freedom in
choosing the origin of the $x$ axis. We set the origin so that $B_1 =
0$. Then, the solution of Eq.\ (\ref{eq:hopsteady2}) is
\begin{eqnarray}
\label{eq:hopsol}
B_o &&= {2m \over 3-m} \cdot {v_s \over W_o} \cr A_o &&= - {p_o \over
W_o} \cdot {6 \over 3-m} \cr A_1 &&= {v_s^2 \over (m-2) g W_o} \cdot
[{2m \over 3-m} (p_o - f_{xx}^o p_o ^m {D \over W_o}) - f_{xx}^o
p_o^m].
\end{eqnarray}
Intuitive solutions for hoppers are that the magnitude of the velocity
increases, and the density decreases as $x$ is increased. Although the
solution Eq.\ (\ref{eq:hopsol}) becomes intuitive one for $0 < m < 3$,
it can also have very different behaviors for other ranges of $m$.

Having obtained the steady state for hoppers, we now proceed to study
kinetic waves. The density and velocity fields are described by Eq.\
(\ref{eq:hoptry}) with one free parameter $p_o$. Consider two regions
of density, where $p_o$ is chosen to be $p_1$ and $p_2$, respectively.
The velocity of the interface $U_{h}$ between the regions, which can
be calculated following the way described in Sec.\ \ref{sec:dvsk}, is
\begin{equation}
U_{h} = {\Delta j \over \Delta \rho} = {\Delta [p_i v_s(p_i) +
v_s{p_i} A_1(p_i) \theta + (p_i B_o(p_i) + v_s(p_i) A_o(p_i)) x
\theta ] \over \Delta [p_i v_s(p_i) + (A_o(p_i) + A_1(p_i) x) \theta
]}.
\end{equation}
We consider small fluctuations of density: $p_1 = p$ and $\delta p =
p_2 - p \ll 1$. The velocity, up to the first order of $\theta$, is
\begin{eqnarray}
\label{eq:hopgroup}
U_{h} \simeq {dj \over d\rho } = && {d \over dp} (p v_s(p)) + {d \over
dp} (A_1(p) v_s(p)) \theta \cr + && {d \over dp} (p B_o(p) + v_s(p)
A_o(p)) x \theta - {d \over dp} (p v_s(p)) {d \over dp} (A_o(p) x +
A_1(p)) \theta.
\end{eqnarray}
The above velocity $U_{h}$ is calculated at position $x$. One should
note that the density at the position is not $p$ but $p + (A_o (p) x +
A_1(p)) \theta $. We now compare the velocity of density waves in
hoppers and tubes. The velocity of density waves in tubes $U_{t}$ at
the above density is
\begin{equation}
U_{t} = {d \over dp} (p v_s(p)) + ({d v_s(p) \over dp} + p {d v_s(p)
\over dp} + p {d^2v_s(p) \over dp^2})(A_o(p) x + A_1(p)) \theta,
\end{equation}
which is give by Eq. \ (\ref{eq:rekvel}).

Therefore, density waves also exist in hoppers, and their velocity is
given by $U_{h}(\theta ,p,x) = U_{t} + C(p,x) \theta,$ where $C(p,x)$
is a complicated function of $x$ and $p$.

\section{Molecular dynamics simulations}
\label{sec:numerical}

\subsection{Interactions between particles}
\label{sec:inter}

We discuss the interaction between the particles used in the MD
simulations of granular flows. The force between two particles $i$ and
$j$, in contact with each other, is the following. Let the coordinate
of the center of particle $i$ ($j$) to be $\vec{R}_i$ ($\vec{R}_j$),
and $\vec{r} = \vec{R}_i - \vec{R}_j$.  In two dimensions, we use a
new coordinate system defined by the two vectors $\hat{n}$ (normal)
and $\hat{s}$ (shear). Here, $\hat{n} = \vec{r} / {\vert \vec{r}
\vert}$, and $\hat{s}$ is defined as rotating $\hat{n}$ clockwise by
$\pi/2$.  The normal component $F_{j \to i}^{n}$ of the force acting
on particle $i$ by $j$ is
\begin{mathletters}
\label{eq:mdforce}
\begin{equation}
\label{eq:fnormal}
F_{j \to i}^{n} = k_n (a_i + a_j - \vert \vec{r} \vert)^{3/2} -
\gamma_n m_e (\vec{v} \cdot \vec{n}),
\end{equation}
where $a_i$ ($a_j$) is the radius of particle $i$ ($j$), $m_i$ ($m_j$)
the mass of particle $i$ ($j$), and $\vec{v} = d\vec{r}/dt.$ The first
term is the Hertzian elastic force, where $k_n$ is the elastic
constant of the material. And, the constant $\gamma_n$ of the second
term is the friction coefficient of a velocity dependent damping term,
$m_e$ is the effective mass, $m_i m_j/(m_i + m_j).$ The shear
component $F_{j \to i}^{s}$ is given by
\begin{equation}
\label{eq:fshear}
F_{j \to i}^{s} = - \gamma_s m_e (\vec {v} \cdot \vec {s}) - {\rm
sign} (\delta s) ~ {\rm min}(k_s \vert \delta s \vert, \mu \vert F_{j
\to i}^n \vert),
$$
\end{equation}
\end{mathletters}
where the first term is a velocity dependent damping term similar to
that of Eq.\ (\ref{eq:fnormal}). The second term is to simulate static
friction, which requires a {\it finite} amount of force ($\mu F_{j \to
i}^{n}$) to break a contact \cite{cs79}. Here, $\mu$ is the friction
coefficient, $\delta s$ the {\it total} shear displacement during a
contact, and $k_s$ the elastic constant of a virtual spring. There are
several studies on granular systems using similar interactions.
However, only a few of them \cite{cs79,bg91,lh93} include static
friction. A particle can also interact with a wall. The force on
particle $i$, in contact with a wall, is given by Eqs.\
(\ref{eq:mdforce}) with $a_j = \infty$ and $m_e = m_i$. A wall is
assumed to be rigid, i.e. it is not affected by the collisions with
particles. Also, the system is under a gravitational field $\vec{g}$.
We do not include the rotation of the particles in present simulation.
A detailed explanation of the interaction is given elsewhere
\cite{lh93}.

\subsection{Density waves in vertical tubes}
\label{sec:nutube}

\subsubsection{Waves without static friction}

We first simulate granular flows without static friction. Thus, we set
$\mu = 0$, and the shear force is only due to the velocity dependent
friction term in Eq.\ (\ref{eq:fshear}). We study the system in two
dimensions. Tubes are modeled by two parallel sidewalls of length $L$,
and separated by distance $W$, between which particles flow (Fig.\
\ref{fig:tubeset}). We use a periodic boundary condition in the
vertical direction. Particles come out of the bottom of the tube are
fed into the top. In order to avoid a haxagonal packing formed by
monodisperse particles, we use polydisperse particles, whose radii are
drawn from the gaussian distribution of mean $0.1$ and width $0.02$.
We initially arrange particles to be equally spaced along the vertical
direction, and calculate the positions and the velocities of the
particles at subsequent steps using a fifth-order predictor-corrector
method.

In Fig.\ \ref{fig:mdtube}, we show the time evolutions of the
densities and the velocities of the particles with $W=1$, $L=15$ and
the number of particles $N = 225$. The density plots are made as
follows. We divide the tube into several regions (bins) of equal
height (typically, $5$ times particle diameter), and count the number
of particles $n_{i}$ in bin $i$. We set the grayscale of each bin to
be proportional to $n_{i}$. We choose white for $n_i = d_l$, and black
for $n_i = d_u$, where $d_l$ ($d_u$) is the lower (upper) bound for
$n_{i}$. If $n_{i}$ is smaller than $d_l$ or larger than $d_u$, the
grayscale is chosen to be white or black, respectively. In Fig.\
\ref{fig:mdtube}(a), we use $d_l = 0$ and $d_u =30$. The density
field at a given time step is plotted as one horizontal line, where
boxes of different grayscale represents bins of the tube. Here, the
leftmost box corresponds to the bin at the bottom of the tube. The
velocity plot is made using the same procedure as above, except the
grayscale is proportional to the average vertical velocity $v_i$ in
bin $i$. In order to enhance the contrast, we subtract the center of
mass velocity from $v_i$. We choose white for $v_l$ and black for
$-v_l$. Here, we set $v_l = 60$. The time step is chosen to be $5.0
\times 10.^{-5}$ The time interval between the successive raws in the
density and velocity plots is $100$ iterations. The parameters for the
simulation are $k_n = 1.0 \times 10^{6},$ $k_s = 1.0 \times 10^{4}$
and $\gamma_n = \gamma_s = 500$ between the particles. Between the
particles and the walls, we use $k_n = 5.0 \times 10^{6}$ in order not
to allow the particles go through the walls, while the other
parameters are kept to be the same.

In Fig.\ \ref{fig:mdtube}, one can see a region of high density is
being formed from the homogeneous system \cite{l93}. Also, a high
density region may be split into two, or two regions may merge to form
one region.  However, for the most of time, these density fluctuations
just travel with almost constant velocity. These traveling density
patterns are first observed in simulations by P\"{o}schel \cite{p92}.
Comparing Fig.\ \ref{fig:mdtube}(a) and (b), one can notice
correlations between the density $n_i$ and the velocity $v_i$. The
particles seem to travel slower (faster) in high (low) density
regions, which is very similar to traffic flows. This correlation is
one hint that the density waves may be of kinetic nature. In order to
systematically study the correlation, we measure the density
dependence of the flux in a steady state. We choose a bin, and we
calculate the friction and gravity force acting on the bin.  Since we
want to measure the flux in a steady state (the total force is zero),
we require the total force (sum of the friction and the gravity) is
smaller than $r$ times the gravity, and we discard bins if the
requirement is not fulfilled. For bin $i$ in a steady state, we
measure the total flux, defined as $j_i = v_i n_i$. We calculate the
density $n_i$ dependence of the average flux $j_i$ for a system of
fixed total number of particles $N$. Here, the averages are taken over
time. We measure the flux-density curve for several different values
of $r$ ranging from $0.1$ to $\infty$.  The results are not very
sensitive to the values of $r$, when we study the systems in a steady
state. We set $r=1.0$ from now on. We also measure the curve for
several values of $N=150, 225, 280, 337$. For large $N$, we have
accurate estimate of $j_i$ for large values of $n_i$, but poor one for
small $n_i$.  The situation is opposite for the systems of small $N$.
For the intermediate values of $n_i$, however, all the systems give
good estimates, which agree with each other. In Fig.\
\ref{fig:fdtube}, we show the flux-density curve averaged over the
four values of $N$. The fact that we have a well defined flux-density
curve suggest the system is in a steady state, which implies that
kinetic wave is sufficient for the description of the evolutions of
the system. Furthermore, the curve resembles that of a traffic flow,
which is considered to be one of the typical examples of kinetic waves
\cite{lw55}.

One additional evidence that the density waves are of kinetic nature
is the dependence of the velocity of the waves on the average density.
The velocities of kinetic waves are given by Eq.\ (\ref{eq:kvel}),
which are the slope of the flux-density curve, for small density
fluctuations. From Fig.\ \ref{fig:fdtube}, we expect that the velocity
is a large negative value for a small density, approaches to zero, and
becomes a large positive number as the density is increased.  We
directly measure the velocities of the waves from the slopes of high
density regions like the one shown in Fig.\ \ref{fig:mdtube}. In
Table.\ \ref{table:vdense}, we show the average velocities for several
values of $N$, where the average density $< n_{i} >$ is given by $N /
L$. In the table, one note the velocity is negative ($-41$) for small
$N$, and is increased to positive ($113$) for large $N$, which is
exactly the way predicted by the theory of kinetic wave.  Furthermore,
the measured velocities are consistent with the local slopes of the
flux-density curve Fig.\ \ref{fig:fdtube}, although the slopes can not
be accurately determined due to the large error bars. Based on the
above evidences, as well as the theoretical argument given in Sec.\
\ref{sec:thtube}, we conclude the above density waves found in
vertical tubes are kinetic waves.

We also want to discuss the origin of fluctuations in the system. As
shown above, density fluctuations (waves) are formed from an uniform
density system. But kinetic waves, as discussed in Sec.
\ref{sec:dvsk}, can not create fluctuations. Also, some of the waves
are split into two waves, which can not be described by the evolutions
of kinetic waves alone. There must be some sources of fluctuations or
``noises'' in the system.  Since the system is deterministic, one
might think the system can not have noises. The ``noises'' come from
the fact that the equations of motion Eq.\ (\ref{eq:motion}) as well
as the form of the friction force are relations between {\it averaged}
quantities. The fluctuations around their averaged values, especially
relevant in small scale descriptions of systems, are identified as
``noises''.

\subsubsection{Waves with static friction}

We next study the flows through vertical tubes {\it with} static
friction. As discussed in Sec.\ \ref{sec:thtube}, we expect two types
of behaviors depending on the friction coefficient $\mu$. If $\mu >
\rho g W / P_{s}$, the system can not move, and stays as a static
column.  Otherwise, the particles start move, and forms periodic
stagnant zones whose separation ${\cal D}(p)$ is determined by Eq. \
\ref{eq:tubesep}.  The setup we used in the MD simulations is exactly
the same as the previous one (Fig.\ \ref{fig:tubeset}), where static
friction is introduced by choosing a non-zero $\mu$. Between the
particles, we use $\mu _{PP} = 0.5$. We also set the shear friction
coefficient $\gamma _{s}$ to be zero. We do simulations with $W=1$ and
$L=15$, starting with $N=225$ particles. All the other parameters are
kept to be the same as above.  In the simulations, we find three types
of behaviors depending on the friction coefficient between the walls
and the particles $\mu _{WP}$: (1) If $\mu _{WP} \ge 1.0$, the
particles can move initially, but they eventually form static
column(s). (2) On the other hand, if $\mu _{WP} \le 0.6$, the
particles constantly increase their speeds, and do not go into a
steady state until the end of simulations (50 000 iterations). (3) In
the intermediate regime, $0.6 \le \mu _{WP} \le 1.0$, we find
steady states, where density waves travels with almost constant
velocities. The static structures found in the first regime are
precisely what is expected from the theory. In the second regime,
however, the systems do not reach steady states, in contrast to the
theory. One possibility is that the time needed to reach a steady
state is larger than the simulation time. To check this possibility we
do longer simulations of the system with $\mu _{WP} = 0.6$, which does
not reach a steady state in $50 000$ iterations. We find the system
{\it does} reach a steady state at around $100 000$ iterations.

In the third regime, the simulations with $L = 15$ do not show the
expected periodic stagnant zones but traveling density fluctuations.
For example, we show the density evolutions in a tube of $L = 15$ in
Fig.\ \ref{fig:mdstube}(a), where $\mu _{WP} = 0.9$ and the time
interval between the successive raws are $0.0025$. A region of high
density travels with almost constant velocity of $-86 \pm 13$. There
are few possibilities to understand the discrepancy. First of all,
there can be a problem of commensurability. In general, the length of
the tube $L$ is not an integer multiple of the stagnant zone
separation ${\cal D}(p)$. Thus, the distance between the clogged zone
can not all be ${\cal D}(p)$ due to the periodic boundary condition.
It can be shown, following the argument in Sec.\ \ref{sec:thtube},
that the distances between the stagnant zones are all ${\cal D}(p)$
except one, which is smaller than ${\cal D}(p)$. This configuration,
however, can not be a steady state. The stagnant zone just below the
small separation becomes unstable, since the incoming velocity of the
particles are smaller than $v_c(p)$. These particles travel further
down to reach $v_c(p)$, then form a stagnant zone. In effect, the
stagnant zones travels down, if $L$ is not an integer multiple of
${\cal D}(p)$.  There also can be problems of ``noises.'' As discussed
before, there are noises to the continuum description we have been
using. Due to these noises, the separations between the zone are not
all equal, but fluctuate around ${\cal D}(p)$. Furthermore, the noise
also cause the stagnant zones to move around.  Considering the above
possibilities, we expect that several stagnant zones in a long tube
drift downwards, whose separations fluctuate around ${\cal D}(p)$. In
order to check the possibilities, we study density waves for several
values of $L$, keeping the average density constant. The evolution of
density field for $L = 30$ is shown in Fig.\ \ref{fig:mdstube}(b). One
can see a region of high density is traveling downwards with the
velocity around $-25$. The results for both $L = 15$ and $L = 30$ are
consistent with the prediction. Since there are only one region of
high density, we expect ${\cal D}(p)$ is larger than $L$. We then
expect the velocities of the density waves are always negative, and
their magnitude decreases as one increases $L$, thereby approaching
${\cal D}(p)$. The situation for $L = 45$ is quite different, as shown
in Fig.\ \ref{fig:mdstube}(c). The high density region seems to travel
upwards, and the velocity seems to fluctuate more. These two features
are also found in a few additional runs we study for $L = 45$. The
fluctuations can be caused by the ``noise,'' but the trend of moving
upwards can not be explained. We also studied the system with $\mu
_{WP}= 0.7$ and $0.8$, and find essentially the same.

\subsubsection{Comparison with the experiments}

We now discuss the results of the experiments with vertical tubes, and
compare them with the theoretical and simulational results obtained
above. We first discuss the experiments. Density waves in vertical
tubes are first found in the experiments by P\"{o}schel \cite{p92}. He
found two types of waves: (1) Regions of large densities occur at
random positions, and they travel with non-zero velocity. (2) The
separations between the regions of large densities seem to be about
the same, and they fluctuate around certain positions \cite{p93}. The
conditions needed to obtain each types of waves are unknown. The flows
are also studied in vacuum, and surprisingly, the density waves
disappear \cite{b93,h93}. It is still not understood {\it how} air
affects the formation of density waves. Here, we discuss one of the
possible mechanisms. The conditions for forming kinetic waves, as
discussed in Sec.\ \ref{sec:thtube}, are fairly simple. One important
condition is that the friction force should balance the driving force,
which in this case, gravity. It is possible that the friction force by
the sidewalls is too small to balance gravity. Consider a block of
particles falling down in a tube. If the density of the block is large
enough, air can not easily pass through the block, and the air
pressure just behind the block can be smaller than that of the front.
The pressure difference gives rise a force to slow down the block,
which acts as an additional friction force. Since the pressure
difference is expect to increase by increasing the density and
increasing the velocity of the block, the friction force by air can
balance gravity at high velocity, and the balance produces kinetic
waves. As one can see, this argument is largely speculative, and
should be checked by careful experiments. Especially, it should be
check whether there is pressure difference between the front and the
behind of a moving granular blocks of high density, and whether the
difference is enough to balance the gravity.

Assuming the above mechanism to be hold, how can one understand the
two types of waves found in the experiments? The two waves are readily
compared with the predicted kinetic waves with and without static
friction. If ${\cal D}(p)$ is larger than the length of the tube, we
do not expect to see any stagnant zone (for open boundary condition).
Therefore, we effectively see the system without static friction,
which produces traveling kinetic waves. However, if ${\cal D}(p)$ is
much smaller than the tube length, we {\it do} see periodic stagnant
zones. The distance ${\cal D}(p)$ is a decreasing function of the
friction (both by the sidewalls and air), and it is possible that the
tube which produces the periodic zones have larger friction than
others. This possibility again is speculative, and should also be
checked by more controlled experiments. For example, the friction can
be changed by using the sidewalls of varying surface roughness. If air
is responsible for an additional friction, one can also change the
friction by controlling the pressure of air.

\subsubsection{Dynamic waves}

We conclude this section by discussing dynamic waves in granular flows
through tubes. As discussed in Sec.\ \ref{sec:dynamic}, we expect the
granular flows become less stable under density fluctuations for
smaller coefficient of restitution $e$. Since $e$ is small for large
normal damping coefficient $\gamma _{n}$, we expect the system to be
less stable for large $\gamma _{n}$. We simulate the system with
several values of $\gamma _{n} = 5 \times 10^{2}, 1 \times 10^{3}, 2
\times 10^{3}$, $3 \times 10^{3}$, and study the stability. In Fig.\
\ref{fig:dtube}, we show the density fields of tube with $L = 15$ and
$N = 225$ for several values of $\gamma _{n}$. Here, the time
intervals between the successive raws are $0.0025$. We do not find any
sign of dynamic waves, even for longer simulations. We also repeat the
simulations with higher density $N = 337$, and find that the results
are essentially the same. The coefficient of restitution $e$ depends
on the relative velocity between the colliding particles for the
Hertzian contact force. Considering the scale of velocities found in
tubes, $e$ is roughly estimated to be $0.1$ for $\gamma _{n} = 3
\times 10.^{3}$ It is possible that the dynamic instability occur at
even larger values of $\gamma _{n}$. The limitation of the constant
timestep type algorithm we are using is that we have to decrease the
timestep to prevent a numerical instability, if we want to increase
$\gamma _{n}$. For example, the timestep is chosen to be $5 \times
10^{-5}$ for $\gamma _{n} = 5 \times 10^{2}$, and $1 \times 10^{-5}$
for $\gamma _{n} = 3 \times 10^{3}$. Therefore, we can not simulate
systems of arbitrarily large $\gamma _{n}$. One have to use event
driven type algorithms (for example, Ref.\ \cite{hl90}) to overcome
this limitation.

\subsection{Density waves in hoppers}
\label{sec:nuhop}

In Sec.\ \ref{sec:thhop}, we presented a theory which predicts the
existence of kinetic waves in granular flows through hoppers. In that
section, the velocity of the kinetic wave in hoppers of opening angle
$2 \theta$ is shown, up to the first order of $\theta$, to be $U
_{h}(\theta ,p,x) = U _{t}(p) + C(p,x) \theta$. Here, $p$ is the
packing fraction and $x$ the position, and $U _{t}(p)$ the velocity of
kinetic waves in tubes, and $C(p,x)$ is a complicated function of $p$
and $x$. We thus expect $U _{h}$ for a hopper of small $\theta $ is
not very different from that of tubes. The conditions for the
existence of dynamic waves are too difficult to be obtained from the
theory.

We study hoppers of length $L = 15$, bottom width $W _{o}= 1.0$, and
several values of the opening angle $2 \theta$. We apply a periodic
boundary condition in the vertical direction. The particles come out
of the bottom are again fed into the top. The boundary condition,
which is not natural for the hopper geometry, can introduce some
artifacts to the system. We later check the results by comparing with
those from the open boundary condition. The main reason for using the
periodic boundary condition is that we can simulate the systems for
longer time, which in effect study the systems of larger sizes without
actually increasing the length $L$ (and the number of particles). We
initially arrange particles as a square lattice in a hopper. The
lattice constant is $0.1$, the average radius of the particles. The
average number of particles per unit area is very close to $25$, the
maximum density allowed for the square packing. The randomness in the
initial configuration is introduced by the polydispersity of the
particles. The interaction parameters are chosen to be exactly the
same as those for tubes.

We first study the system without static friction. The static friction
coefficient $\mu$ is chosen to be $0$, and the velocity dependent
shear friction term $\gamma _s$ to be $500$. In Fig.\ \ref{fig:mdhop},
we show the time evolutions of the density and the velocity fields in
hoppers. The plots are made in the same way as Fig.\ \ref{fig:mdtube}
except the fact that the density is obtained by dividing the number of
particles in a bin by the area of the bin, which is not constant for a
hopper. The grayscales are chosen to show clear contrast between the
low and high density (velocity) regions by controlling $d_l$ and $d_u$
($v_l$).

In Fig.\ \ref{fig:mdhop}(a) and (b), we show the density and the
velocity fields for a tube ($\theta = 0$) in order to serve as a
reference to compare with those for hoppers. The time interval between
the successive lines are $0.005$.  In Fig.\ \ref{fig:mdhop}(c) and
(d), the fields for a small angle hopper ($\theta = 1^{\circ}$) are
shown.  In the density plot (Fig.\ \ref{fig:mdhop}(c)), there are two
waves in the hopper, one travels upwards and the other downwards.
Comparing with the velocity plot (Fig.\ \ref{fig:mdhop}(d)), the
densities of the upward wave are strongly correlated to the
velocities, while there seems to be no such correlation for the
downward waves. The correlations found in the upward waves are
qualitatively the same as that of kinetic wave, i.e., the particles
travel slow (fast) in the regions of high (low) density.  It is quite
possible that the upward wave is of kinetic nature. Also, the
velocities of kinetic waves in hoppers with small opening angle are
expected to be very close to that of tubes.  The fact that the
velocity of the upward waves ($113 \pm 4$) are very close to that of
the tube ($93 \pm 10$) is one other support that the wave is kinetic
\cite{note1}. We now consider the waves travel downwards. The waves,
not only show no correlations between the density and the velocity
fields, but also travels in the opposite direction to the kinetic
waves in the tube (Fig.\ \ref{fig:mdhop}(a)), which suggests that the
waves are not probably kinetic.  Furthermore, the density contrasts of
the waves are constantly increasing, which is not possible for kinetic
waves. The increments can be more clearly seen in hoppers of larger
opening angle as shown in (e)-(f) ($\theta = 4^{\circ}$). The
downward waves initially coexist with the kinetic waves, but
eventually dominate the system. The above facts are still true for the
largest $\theta =10^{\circ}$ we study. The properties of the downward
waves listed above suggest that the downward waves are of dynamic
nature. According to the argument used for the dynamic waves in tubes,
dynamic waves will be more easily formed using non-elastic particles.
We check this possibility by studying a hopper of $\theta = 1^{\circ}$
for different values of $\gamma _{n}$. In Fig.\ \ref{fig:dhopper}, we
show the density fields for $\gamma _{n} = 1 \times 10^{3}$ and $2
\times 10^{3}$. Comparing with that of $\gamma = 5 \times 10^{2}$
(Fig.\ \ref{fig:mdhop})(c), one can see the waves for larger $\gamma
_{n}$ are indeed formed earlier, and their intensities (density
contrast) are larger. This add one more support that the downward
waves are dynamic. We also measure the velocities of the dynamic waves
for several values of $\theta$. As shown in Table \ref{table:vhop},
the magnitude of the velocity is decreased as $\theta$ is increased.
Following the trend, it is quite possible that the velocity becomes
zero for finite $\theta$, and change its sign.

We now study hoppers with the open boundary condition. In Fig.\
\ref{fig:mdohop}(a) and (b), we show the density and the velocity
fields for a hopper of $\theta = 1^{\circ}$ and $L = 30$. Here, we
turn off static friction ($\mu = 0$), and we set $\gamma _{s} = 5
\times 10^{2}$. One can hardly see any fluctuations in the density
field, but one can see some traveling patterns in the velocity field.
We now turn on static friction. We set $\gamma _{s} = 0$ and $\mu \neq
0$. The density and the velocity fields with $\mu = 0.5$ are shown in
Fig.\ \ref{fig:mdohop}(c) and (d). One can now see upward traveling
density waves. Furthermore, the density field have strong correlations
with the velocity field, suggesting that the waves are kinetic. Even
though kinetic waves are present in the system independent of static
friction, the amplitudes of the waves are too small to be visible
without it. Static friction provides an effective mechanism for
creating large density fluctuations. Also, the density waves are not
present in the hoppers of small $L$. The absence of density waves can
be caused by the fact that the particles need to travel certain
distances to reach a steady state around which the density waves can
only be seen. All these results remain valid for larger values of
$\theta$ we have studied. For example, we show the case of $\theta =
5^{\circ}$ in Fig.\ \ref{fig:mdohop}(e)-(f).

We also search for the dynamic waves in hoppers with the open boundary
condition. We simulate the system with several values of $\gamma
_{n}$, and check the instability of creating dynamic waves. For the
range of $\gamma _{n}$ we have studied ($5 \times 10^{2} \sim 3 \times
10^{3}$), we do not find the instability. Again, the simulations for
larger values of $\gamma _{s}$ are limited due to the algorithm we are
using. Even though we do not rule out dynamic waves in hoppers in
general, the dynamic instability found above seems to be an artifact
of the periodic boundary condition.

We now compare the results with the experiments. Baxter {\it et al}
found density waves in a hopper only if a certain amount of rough
sands is mixed with smooth sands \cite{bbfj89}. The role of the rough
sands is not clearly established. It is possible that the rough sands
forms local ``arches'', thereby increasing density fluctuations. In
that case, the role played by the rough sands is the same as that of
static friction, which provides density fluctuations to maintain
kinetic waves. Also this is consistent with the fact that the
amplitudes of the density waves are a smooth function of the fraction
of the rough sands, which suggests that the density waves in the
experiments are kinetic. The suggestion, of course, has to be checked
by experiments, for example, by measuring the correlations between the
density and the velocity fields.

\section{Discussion}
\label{sec:dis}

We have presented theoretical and numerical evidences that the density
waves found in the simulations of P\"{o}schel is of kinetic nature.
However, the density waves found in the experiments are not fully
understood. The first and foremost problem is to find the form of the
friction force. In the MD simulations, the friction force is generated
by the collisions of particles to the sidewalls. In experiments, the
collisional friction force seems to be too small compared to gravity,
and another friction force related to air gives dominant contribution.
We proposed a mechanism how air can generates a friction force, which
should be checked by experiments. We want to emphasize that since the
existence of kinetic wave is not strongly dependent on the details of
the friction force, the density waves found in the experiments are
very likely to be kinetic waves.

We find conditions for the existence of dynamic waves in tube. The
conditions depend on details of the force, which we do not know. We
numerically search for dynamic waves by increasing $\gamma _{n}$,
since we expect dynamic waves can be more easily created for inelastic
particles. Even for the largest $\gamma _{n}$ we studied, we can not
find the dynamic waves. It is possible that dynamic waves occur for
even larger values of $\gamma _{n}$. We are not able to check the
possibility, since the algorithm we are currently using becomes very
inefficient for larger values of $\gamma _{n}$.

We also have present theoretical and numerical evidences that there
are kinetic waves in hoppers. Especially, the kinetic waves in hoppers
with the open boundary condition are visible {\it only} with static
friction. This can be readily compared with the fact that one need
finite fraction of rough sands to observe the density waves in the
experiments. Here, the role of rough sands can be creating large
density fluctuations to maintain kinetic waves just like static
friction. The suggestion that the density waves in the experiments of
hoppers are kinetic waves should be checked, for example, by studying
the correlation between the density and the velocity fields.

We are not able to do a linear stability analysis for hoppers, due to
the complicated density and velocity fields in the steady state. In
the simulations with the peridic boundary condition, we find another
density waves, which we believe to be dynamic on a few evidences.
Since we do not find the dynamic waves with the open boundary
conditions, we think the above dynamic wave is an artifact of the
boundary condition. However, we do not rule out the dynamic waves in
hoppers, especially for large values of $\gamma _{n}$.

I thank Michael Leibig, Hans Herrmann and Thorsten P\"{o}schel for
many useful discussions.

\begin{figure}
\caption{The time evolution of the interface between the uniform-flows
with density $\rho _a$ and $\rho _b$.}
\label{fig:inter}
\end{figure}

\begin{figure}
\caption{Time evolution of density fluctuation with the force given by
Eq.\ (2): (a) Dynamic waves are decaying, while the kinetic waves
survives ($\mu = 2, \nu = -1$), (b) Dynamic waves are growing, and
take over kinetic waves ($\mu = -2, \nu = -1$).}
\label{fig:dvsk}
\end{figure}

\begin{figure}
\caption{(a) Hopper with the opening angle $2 \theta$. (b) The friction
force per unit length on the wall is sum of $\tau _{xy} \cos \theta$
(left) and $\tau _{xx} \cos \theta \sin \theta$ (right).}
\label{fig:hopper}
\end{figure}

\begin{figure}
\caption{Vertical tubes in two dimension with width $W$ and length $L$
under gravity. We apply a periodic boundary condition in the vertical
direction.}
\label{fig:tubeset}
\end{figure}

\begin{figure}
\caption{Time evolutions of (a) density and (b) velocity fields of
particles in a tube of $W = 1, H = 15$. Fields at a given time is
shown as a horizontal line of boxes. The grayscale of each box is
proportional to the density and the velocity in that region of the
tube. Regions of high density are formed, and travel with almost
constant velocity.}
\label{fig:mdtube}
\end{figure}

\begin{figure}
\caption{The flux-density curve for a tube of $W=1, H=15$ averaged
over time and different values of $N$. The parabolic shaped curve
resembles that found in a traffic jam.}
\label{fig:fdtube}
\end{figure}

\begin{figure}
\caption{The evolution of density fields in tubes with $\mu = 0.5$ and
(a) $L = 15$, (b) $30$ and (c) $45$. High density regions travel
downwards in (a) and (b), but travel upwards with larger fluctuations
in (c).}
\label{fig:mdstube}
\end{figure}

\begin{figure}
\caption{Density fields for tube with $L = 15$ and $N = 225$ for
several values of $\gamma _{n}$: (a) $\gamma _{n} = 1 \times 10^{3}$,
(b) $2 \times 10^{3}$ and (c) $3 \times 10^{3}$. There is no sign of
dynamic instability.}
\label{fig:dtube}
\end{figure}

\begin{figure}
\caption{The density and velocity fields for hoppers with different
opening angle. (a)-(b) $\theta = 0^{\circ}$, (c)-(d) $1^{\circ}$,
(e)-(f)  $4^{\circ}$. For small $\theta$, there are two waves traveling
in opposite directions, where the downward waves eventually dominate.}
\label{fig:mdhop}
\end{figure}

\begin{figure}
\caption{The density fields for a hopper of $\theta = 1^{\circ}$ with
(a) $\gamma _{n} = 1 \times 10^{3}$ and (b) $2 \times 10^{3}$. The
intensities (density contrast) of waves are larger for larger $\gamma
_{n}$.}
\label{fig:dhopper}
\end{figure}

\begin{figure}
\caption{The density and velocity fields for a hopper with the open
boundary condition. In (a)-(b) the fields for a hopper of $\theta =
1^{\circ}$ without static frcition, (c)-(d) for $\theta = 1^{\circ}$
with static friction, and (e)-(f) for $\theta = 5^{\circ}$ with static
friction are shown.}
\label{fig:mdohop}
\end{figure}

\begin{table}
\caption{The velocity of kinetic waves in a tube for several values of
the average density}
\label{table:vdense}
\begin{tabular}{cdd}
N & $< n_{i} >$ & Velocity of kinetic wave \\ \hline
150 & 10.0 & -41.0 $\pm$ 2.0 \\
225 & 15.0 & 5.0 $\pm$ 9.0 \\
280 & 18.7 & 12.0 $\pm$ 11.0 \\
337 & 22.5 & 113.0 $\pm$ 4.0
\end{tabular}
\end{table}

\begin{table}
\caption{The velocity of dynamic waves in a hopper for several values
of $\theta$.}
\label{table:vhop}
\begin{tabular}{cc}
$\theta$ (in degrees) & Velocity of dynamic wave \\
\hline
1.0 & -49 $\pm$ 3 \\
2.0 & -40 $\pm$ 5 \\
4.0 & -26 $\pm$ 4 \\
6.0 & -21 $\pm$ 5 \\
8.0 & -18 $\pm$ 4
\end{tabular}
\end{table}


\begin{references}
\bibitem{note1} The vertical size of pixels in Fig.\ 9(a) and (b) is
twice as large as compared with Fig.\ 9(c) - (f). The velocity of
upward waves in Fig.\ 9(a) is comparable to that of Fig.\ 9(c).
\end{references}
\end{document}